\def\cF{{\cal F}}

\def\cM{{\cal M}}
\def\cN{{\cal N}}
\def\cO{{\cal O}}

\newcommand{\vvv}{\mathrm{v}}

\def\im{{\rm i}}

\def\cC{{\cal C}}

\newcommand{\e}{\mathrm{e}}

\newcommand{\be}{\begin{equation}}
\newcommand{\ee}{\end{equation}}
\newcommand{\bea}{\begin{eqnarray}}
\newcommand{\eea}{\end{eqnarray}}

\newcommand{\ominf}{\omega_{\scriptscriptstyle \infty}}

\newcommand{\?}{\;\!}

\newcommand{\Iprod}[2]{\langle {#1}, {#2} \rangle}

\newcommand{\qv}{\mathcal{R}}
\newcommand{\sv}{\mathcal{I}}

\def\cJ{J}

\def\sl{\frak{sl}}

\def\cC{{\cal C}}
\def\cH{{\cal H}}

\documentclass{PoS}

\usepackage{amsmath}

\title{Holographic duals of refined partition functions}

\ShortTitle{Holographic duals of refined partition functions}

\author{\speaker{Chiara Toldo}\thanks{Contribution based on  \cite{Toldo:2017qsh,Hristov:2018spe,Hristov:2019mqp}.}\\

        Centre de Physique Theorique, Ecole Polytechnique, CNRS, 91128 Palaiseau Cedex, France\\
        
        and \\
       
        Institut de Physique Theorique, Universite Paris Saclay, CEA, CNRS, Orme des Merisiers, 91191 Gif-sur-Yvette Cedex, France \\
        
        and \\
       
        Institute for Theoretical Physics, University of Amsterdam, Science Park 904, Postbus 94485, 1090 GL, Amsterdam, The Netherlands \\
        
        E-mail: \email{chiara.toldo@polytechnique.edu, c.toldo@uva.nl}}

\abstract{Recent years have witnessed lots of progress in the computation of supersymmetric partition functions of SCFTs on curved manifolds via localization. The twisted partition function on product manifolds of the form $S^1 \times \Sigma_g$, where $\Sigma_g$ is a two-dimensional Riemann surface, is of particular relevance due to its role in the microstate counting for magnetic static AdS$_4$ black holes realizing the topological twist. We present here supergravity solutions having as conformal boundary more general 3d manifolds. We first focus on solutions (AdS-Taub-NUT and AdS-Taub-Bolt) having as boundary a circle bundle over $\Sigma_g$, showing the matching of their on-shell action with the large $N$ limit of the partition function of the dual CFT. We then discuss some recent results for a challenging example, which involves the refinement by angular momentum. The gravitational backgrounds in this case are rotating supersymmetric AdS$_4$ black holes. We show how to construct two different classes of such solutions in theories of supergravity with uplift in M-theory, and comment on the current status of their entropy counting in the dual CFT.

}

\FullConference{Corfu Summer Institute 2019 "School and Workshops on Elementary Particle Physics and Gravity" (CORFU2019)\\
		31 August - 25 September 2019\\
		Corfu, Greece}

\begin{document}

\include{tableofcontents}

\section{Introduction}

Black holes are often seen as a theoretical laboratory of a theory of quantum gravity: their entropy provides us invaluable quantitative information about the degrees of freedom of such a theory. In this regard, the AdS/CFT correspondence provides a natural setting for the statistical interpretation of black hole entropy in terms of microstates of the dual field theory. One  of  the  great  successes  of  string  theory  is  the  microscopic explanation of the entropy of certain classes of asymptotically flat black holes \cite{Strominger:1996sh,Maldacena:1997de}. Quite recently the microstate counting was extended to classes of Anti-de Sitter black holes: the entropy of four-dimensional static supersymmetric AdS black holes was obtained \cite{Benini:2015eyy} from the leading contribution to the Witten index of the boundary ABJM \cite{Aharony:2008ug} field theory. The counting procedure involves the computation of the partition function of the dual field theory on a rigid background of the form $S^1 \times \Sigma_g$, where $\Sigma_g$ is a two-dimensional Riemann surface of genus $g$, with magnetic flux through $\Sigma_g$. This opened the way to a number new developments, also involving higher dimensional setups, which are reviewed for instance in \cite{Zaffaroni:2019dhb}. These results have fueled new technical advances both in supergravity and in the exact computation of partition functions by means of supersymmetric localization. 

In this paper we would like to review recent results regarding more general supersymmetric partition functions of 3d $\mathcal{N}=2$ SCFTs and their holographic duals, focusing on two cases. The first concerns the partition function of $\mathcal{N}=2$ SCFTs on $U(1)$ p-bundles over $\Sigma_g$, recently computed in \cite{Closset:2017zgf} via localization using a three-dimensional uplift of the 2d $A$-model. The 3-manifold, dubbed $\mathcal{M}_{g,p}$, is parametrized  by two integers $g,p$ and is described by a metric of the form
\begin{equation}\label{Mgp}
ds^2 =  (d \psi +a)^2 + d\Omega_{\kappa}^2 
 \qquad -\frac{1}{4 \pi} \int_{\Sigma_g} da =p \,. \end{equation}
For particular choices of $p,g$, the computation reduces to known cases. For instance, the choice $g=0, p=1$ corresponds to the partition function on $S^3$, satisfying monotonicity theorems \cite{Jafferis:2011zi} and related to the  (finite  terms)  in  the entanglement entropy for a disk region in flat space in the dual field theory \cite{Casini:2011kv}. Upon setting $g=0, p>1$ one recovers the partition function of 3d theories on Lens space (i.e. \cite{Alday:2012au}), while setting $p=0$ we obtain the partition function used for the black hole microstate counting in \cite{Benini:2015eyy,Benini:2015noa}. The importance of this framework stems from the fact that it allows a unified description of these different objects, making connections between the objects used to compute them. Gravity duals having $\mathcal{M}_{g,p}$ as conformal boundary fall into the class of "NUT" and "Bolt" solutions found in Euclidean minimal gauged supergravity \cite{Martelli:2012sz,Toldo:2017qsh}, and their description be object of Sec. \ref{nuts}.

The twisted supersymmetric partition function admits also a refinement by angular momentum, introduced in \cite{Benini:2015noa} and later developed in \cite{Closset:2018ghr}, characterized by the addition of a fugacity for the angular momentum, achieved by considering the background metric
\begin{equation} \label{refj_intro}
ds^2 = d\theta^2 +f(\theta)(d \phi -\varsigma dt)^2 + dt^2
\end{equation}
where $\varsigma$ is a constant parameter related to the fugacity for the angular momentum. Supersymmetric solutions having this (Euclidean) boundary are rotating AdS$_4$ black holes \cite{Klemm:2011xw,Hristov:2018spe}. The superconformal index also admits a refinement by angular momentum \cite{Kim:2009wb,Imamura:2011su,Kapustin:2011jm}, and its value is related to the entropy of supersymmetric rotating black holes with zero R-symmetry flux \cite{Kostelecky:1995ei,Cvetic:2005zi,Hristov:2019mqp}. The refinement by angular momentum is of particular importance, since the black holes in our universe are often spinning close to extremality, and supersymmetry in AdS$_4$ is compatible with the presence of angular momentum. In Sec. \ref{rotating} we review the construction of two broad classes of these BPS black hole solutions, found in abelian $\mathcal{N}=2$ gauged supergravity coupled to vector multiplets.

This short paper is aimed at giving a summary of the main results of \cite{Toldo:2017qsh,Hristov:2018spe,Hristov:2019mqp}, regarding holographic duals to the two setups described above. In other words, we will consider the task of "filling" boundary geometries of the form \eqref{Mgp} and \eqref{refj_intro} (along with the corresponding values for the other background fields) with regular bulk solutions. In the first case, we will work directly in Euclidean signature, avoiding the closed timelike curves which arise in their NUT charged Lorentzian counterparts. For the second case instead we will present matter-coupled solutions (black holes) with Lorentzian signature. While the scope here is to present a concise review of the salient results, we refer to the original papers for a more exhaustive and detailed derivation.

\section{NUTs and Bolts with $\mathcal{M}_{g,p}$ boundary \label{nuts}}

In this section we describe supersymmetric gravitational backgrounds whose boundary is the manifold $\cM_{g,p}$, a circle bundle over a closed Riemann surface $\Sigma_g$ characterized by metric \eqref{Mgp}. Our starting point is minimal $\cN =2$ four-dimensional gauged supergravity, whose bosonic action reads
\be \label{act}
S = -\frac{1}{16 \pi G_4} \int d^4x \sqrt{g} \left( R +\frac{6}{l^2} -F_{\mu \nu}F^{\mu \nu} \right)\,,
\ee
where $G_4$ is the four-dimensional Newton's constant and $l $ is the AdS radius. We work in Euclidean signature. The gravitino supersymmetry variation is
\be
\delta_{\epsilon} \psi_{\mu} = \left( \partial_{\mu}  -\frac14 \omega^{ab}_{\mu} \gamma_{ab} +\frac12 l^{-1} \gamma_{\mu}  -i l^{-1} A_{\mu} + \frac{i}{4} F_{\rho \nu} \gamma^{\rho \nu} \gamma_{\mu}  \right) \epsilon
\ee
where $\epsilon$ is a Dirac spinor and  $\Gamma_{\mu} $ are the generators of $\text{Cliff}(4,0)$ and so they satisfy $\{\Gamma_a, \Gamma_b \} = 2 g_{ab}$. We follow the conventions of \cite{Martelli:2012sz}. We will restrict our analysis to a set of solutions where $\Sigma_g$ has constant curvature, and to configurations with a real metric. Solutions to the system of equations of motion have the following form\footnote{These solutions were presented in Lorentzian signature in \cite{Chamblin:1998pz}. They are a subset of the more general Plebanski-Damianski ones \cite{Plebanski:1976gy} and are characterized by $SU(2) \times U(1)$ symmetry in the case $\kappa=1$. Solutions with reduced $U(1) \times U(1)$ symmetry were considered for instance in \cite{Martelli:2013aqa} but we do not treat them here.} \be \label{met}
ds^2= \lambda(r)(d\tau +2s f(\theta, \phi))^2 + \frac{dr^2}{\lambda(r)}+ (r^2-s^2) \, d \Omega_{\kappa}^2 
\ee
with
\be
\lambda(r) =\frac{ (r^2-s^2)^2 +(\kappa-4s^2) (r^2+s^2) -2Mr +P^2-Q^2}{r^2-s^2}\,,
\ee
and
\be \label{fff}
f(\theta, \phi) = \left\{ \begin{array}{ccc} \cos \theta d\phi \,\,\, & \text{for} \,\,\,& \kappa=1 \\ -  \theta d\phi \,\,\, &\text{for} \,\,\,&  \kappa=0  \\  - \cosh \theta d\phi \,\,\, &\text{for} \,\,\, & \kappa=-1 \end{array} \right. \qquad d\Omega_{\kappa}^2 = \left\{ \begin{array}{ccc} d\theta^2 + \sin^2\theta d\phi^2 \,\,\, & \text{for} \,\,\,& \kappa=1 \\ d\theta^2 + d\phi^2 \,\,\, &\text{for} \,\,\,&  \kappa=0  \\  d\theta^2 + \sinh^2 \theta d\phi^2 \,\,\, &\text{for} \,\,\, & \kappa=-1 \end{array} \right.
\ee
In this case, $\kappa$ denotes the curvature of $\Sigma_g$: $\kappa=1$ for $S^2$, $\kappa=0$ for $\mathbb{R}^2$ and $\kappa = -1$ for $\mathbb{H}^2$ (which respectively lead, upon a suitable compactification, to $T^2$ and higher genus Riemann surfaces). The gauge field has this form
\be
A_t = \frac{-2 s Q r +P(r^2+s^2)}{r^2-s^2}\,,
\qquad 
A_{\phi} = \left\{ \begin{array}{ccc}\cos \theta \frac{P(r^2+s^2)-2 s \, Q \, r}{r^2-s^2} \,\,\, & \text{for} \,\,\,& \kappa=1 \\ - \theta \frac{P(r^2+s^2)-2 s \, Q \, r}{r^2-s^2} \,\,\, &\text{for} \,\,\,&  \kappa=0  \\  -\cosh \theta \frac{P(r^2+s^2)-2 s \, Q \, r}{r^2-s^2} \,\,\, &\text{for} \,\,\, & \kappa=-1 \end{array} \right. \label{afi}
\ee
In these solutions $M$ is the mass parameter and $r$ is the radial coordinate. $\tau$ parameterizes a circle fibered over a 2-dimensional constant curvature surface $\Sigma_g$ spanned by the coordinates $\theta$ and $\phi$. The fibration is due to the presence of the NUT parameter, which we denote by $s$ because of it parameterizes the "stretching" of the Hopf fiber relative to the 2d base. We set $l=1$ for simplicity in what follows. In the asymptotic limit $r\rightarrow \infty$ the metric approaches
\be \label{boundary}
ds^2 = \frac{1}{r^2} dr^2 +r^2 \left(4s^2 (d \psi + f(\theta, \phi))^2 + d\Omega_{\kappa}^2 \right)\,,
\ee
where we have defined $\psi = \tau/(2s)$.
 In other words, the boundary is a circle bundle over $\Sigma_g$, and in the particular case for which $f(\theta,\phi) = \cos\theta d\phi$ and $\psi $ is periodic with period $\Delta\psi = 4\pi$, the boundary is a squashed 3-sphere with squashing parameterized by $4s^2$.   
 
The bulk solutions we have described in this section can be of the type AdS-Taub-NUT  and AdS-Taub Bolt, depending on the value of the parameters appearing in the warp factor. These solutions are characterized by different topologies, since one of the Killing vectors has a zero-dimensional fixed point set (``nut'') or a two-dimensional one (``bolt''). 
\vspace{2mm}

\vspace{1mm}
\noindent \textbf{NUTs}: For the NUT solution the Killing vector $\partial_{\tau}$ has a fixed point where the $S^2$ has zero radius:
\be \lambda(r=s) =0\,.
\ee This ensures that the Killing vector has a zero-dimensional fixed point. Moreover, absence of Dirac-Misner strings constrain the period of $\tau$ to be
$
\Delta \tau = 4 s \Delta \phi\,,
$
and since $\Delta \phi = 2\pi$, this yields $\Delta \tau = 8s \pi $, or equivalently $\Delta \psi = 4 \pi$ (see formula \eqref{boundary}). The coordinate $\theta$ goes from 0 to $\pi$. Absence of conical singularities at the location of the nut, $r=s$, requires
\be \label{regNUT}
\Delta \tau \lambda'(r=s) =4 \pi \qquad \rightarrow \qquad \lambda'(r=s) = \frac{1}{2s}\,.
\ee
In particular, for NUT solutions the warp factor $\lambda(r)$ has a double root at $ r=s$ and the metric is defined for $r \geq s$.
The point $r=s$ is a NUT-type coordinate singularity and the metric is a smooth metric on $\mathbb{R}^4$ with the origin identified with $r=s$. For $s=1/2$ the squashing vanishes, and the boundary is that of a round $S^3$. If we allow for singularities, we can also take into consideration quotients of the Taub-NUT space, obtaining the AdS-Taub-NUT$/\mathbb{Z}_p$ geometry, which suffers from a conical singularity at the origin.

\vspace{2mm}

\noindent \textbf{Bolts}: For the Bolt solution the Killing vector $\partial_{\tau}$ has a two-dimensional fixed point, so the only condition is that, at a radius $r_b >s$, $ \lambda(r=r_b) =0 $ with $r_b$  a single zero of $\lambda (r)$. For our solution, the fundamental domain for the compactification, $D_g$, is chosen as in \cite{Benini:2016hjo}, so that 
 $
\text{vol}(\Sigma_g) = 4\pi (g-1) \,\,\, \text{for}\,\,\, g > 1
 $ 
 and
$
 \int_{D} d \theta d \phi = 4\pi $ for $g = 1$. Therefore, given that $a =  \theta d\phi$ for the torus, and $a =  \cosh \theta d\phi$ in the higher genus case, we are left with the following conditions
 \be\label{period}
 \Delta \psi = \frac{4 \pi (g-1)}{p} \qquad \text{for } \,\, g>1, \qquad \qquad  \Delta \psi = \frac{4 \pi}{p} \qquad \text{for } \,\, g=1
 \ee
 Moreover, the absence of conical singularities at the location of the bolt requires 
\be \label{regBOLT}
 \frac{r_b^2-s^2}{s \lambda'(r_b)} = \frac{2|g-1|}{p} \equiv  \frac{2}{\mathbf{p}} \,.
\ee
Therefore, a $\mathcal{M}_{g,p}$  boundary with $g>0$ can be filled by the Bolt solution, with topology $\mathcal{O} (-p) \rightarrow \Sigma_g$ and Euclidean time period \eqref{period}. 

The supersymmetry properties of the class of solutions \eqref{met}-\eqref{afi} were analyzed in \cite{AlonsoAlberca:2000cs,Martelli:2011fw,Martelli:2012sz,Toldo:2017qsh}. Two classes of solutions exist, preserving 1/2 and 1/4 of supersymmetry respectively. We focus on the latter, due to their direct connection to the 1/4 BPS black hole solutions. A necessary and sufficient condition for the solutions to be 1/4 BPS is
\be \label{cond_kappa}
P= \mp \frac12 (4s^2-\kappa)\,, \qquad M= \pm 2sQ\,, \qquad Q \,\, \text{unconstrained}\,.
\ee
The Killing spinor was constructed in \cite{Martelli:2012sz,Toldo:2017qsh} has only radial dependence, so that the compactification necessary to obtain a compact Riemann surface is allowed without breaking supersymmetry. Notice that \eqref{cond_kappa} denotes a four-dimensional subspace of BPS solutions, parameterized by $p$ (Chern class of the bundle), $g$ (genus of the $\Sigma_g$ base), $Q$ (electric charge) and $s$ (squashing parameter). However, regularity imposes the additional constraints \eqref{regNUT} for the NUT and \eqref{regBOLT} for the Bolts, so that $Q$ is fixed in terms of the other parameters. The intricate moduli space with different branches of solutions (denoted with Bolt$_{\pm}$) is described in detail in \cite{Martelli:2012sz,Toldo:2017qsh}.

The NUT and Bolts solutions discussed so far uplift locally to 11d, but there are subtleties on the global properties of the uplift that we are now going to discuss. To do so, we notice that the field strength $F$ of the Bolt has a nontrivial magnetic flux through the Bolt surface $\Sigma_g$ at $r = r_0$, computed as   
\begin{eqnarray} \label{fluxbolt}
\int_{\Sigma_g} \frac{F}{2 \pi} & = & - \frac{2s}{r_0(s)^2-s^2} \left[ -2 Q (s) r_0(s) -\frac12(r_0(s)^2+s^2) (4s^2-1) \right]  = \pm \frac{p}{2} -(1-g) \,. \qquad \quad 
\end{eqnarray}
 Its value satisfies $
\int \frac{F}{2 \pi} = \frac{q}{2}
$ where $q = p \mod 2$, which is consistent with  $A$ being a spin$^c$ gauge field.
The 11d uplift ansatz for solutions of (Lorentzian) minimal gauged 4d supergravity was found in \cite{Gauntlett:2007ma}. In order for the 11d metric to be well defined the flux at the Bolt is subject to a further condition. In particular,  depending on the geometry of the internal Sasaki-Einstein 7-manifold $Y_7$, the uplift is possible only for certain values of $p$. When $Y_7$ is a regular Sasaki-Einstein manifold, $Y_7$ is a $U(1)$ bundle over the six-dimensional K\"ahler Einstein manifold $B_6$. Fixing a point on $B_6$, in order for the $U(1)$ bundle to be well-defined at the Bolt itself, the flux should satisfy \cite{Martelli:2012sz}
\be \label{quant_uplift}
\frac{4}{2I} \int_{\Sigma_g} \frac{F}{2 \pi} = m \in \mathbb{Z}\,,
\ee
Where $I$ is the Fano index associated with\footnote{With a slight abuse of notation we will refer to it as the Fano number of the 7d Sasaki-Einstein manifold relative to the base $B_6$.} $B_6$. In particular this means \be
\pm p +2(g-1) = 0 \quad \text{mod} \quad I(Y^7)
\ee
For instance $S^7$ has $I=4$, $V^{5,2}$ has $I=3$, $M^{3,2}$ has $I=1$. Notice that in this latter case Bolt solutions uplift for all values of $p$. The requirement \eqref{quant_uplift} on $p$ and $g$ does not apply Taub-NUT solutions, which have topology $\mathbb{R}^4$ and the gauge field $A$ is globally a one-form on $\mathbb{R}^4$. The NUT solution uplifts without restrictions to eleven dimensions. 

To conclude this section, we record the value of the on-shell action for NUTs and Bolts, computed  via techniques of holographic renormalization  \cite{Skenderis:2002wp}:
 \be \label{FEboltplusminus}
 I_{NUT} = \frac{\pi}{2 G_4} \,, \qquad \quad   I_{Bolt \pm} =\frac{\pi (4(1-g)\mp p)}{8 G_4}\,.
 \ee
The free energy indeed has a remarkably simple form, independent of the squashing parameter $s$. Let us mention that the $g=0$ case was already found in \cite{Martelli:2012sz}. This formula is valid also for the case $p=0$ which deserves a special discussion. Indeed the $p=0$ Euclidean Bolt solutions obtained by Wick-rotating the solution of \cite{Caldarelli:1998hg} reads
\be
ds^2 = U^2(r) dt^2 + U(r)^{-2} dr^2 + r^2 d\Omega_{\kappa}^2\,, \qquad   U^2(r) =  r^2 -1+ \frac{ 1/4-Q^2}{r^2} \,.
\ee
with $A_{t} = \frac{Q}{r}\,, \,\,  A_{\phi} =  \cosh \theta /2$. Such solution caps off smoothly at a finite radial coordinate in the bulk, as long as $Q \neq 0$: in case of vanishing $Q$ the geometry develops an infinite throat characteristic of the AdS$_2$ factor in the near-horizon geometry of an extremal black hole. For $p=0$, $I_{Bolt \pm}$ gives the on-shell action of  $p=0$ Bolts, therefore also in particular for $Q=0$ black hole solutions. Directly related to this, in \cite{Azzurli:2017kxo} it was shown that in the extremal BPS limit the on-shell action of magnetic black holes indeed coincides with their Bekenstein-Hawking entropy. 

Before concluding this section, let us mention that \cite{BenettiGenolini:2019jdz} showed that the on-shell action of any supersymmetric solution to minimal gauged supergravity can be written in terms of topological data, realizing an example of  "localization in the bulk" for classical gravitational partition functions. Formulas \eqref{FEboltplusminus} for $I_{NUT}$ and $I_{Bolt \pm}$ indeed are particular examples of \cite[1.1]{BenettiGenolini:2019jdz}, which displays the separate contributions of NUT ad Bolt type fixed points to the renormalized on-shell action.

\subsection{NUTs and Bolts: comparison with dual field theory}

In this section we compare the gravity result \eqref{FEboltplusminus} with the field theory computation. We focus on the ABJM model for concreteness, commenting on more general models later on. In the supergravity side this means we will consider the uplift on $Y_7=S^7$ first. Different choices of $Y_7$ correspond to different choices of Chern-Simons-matter theory on $\mathcal{M}_{g,p}$.

For the NUT solution the comparison with the field theory result was already performed in \cite{Martelli:2012sz}, by noticing that the NUT on-shell action, using $\frac{1}{G_4} = \frac{2 \sqrt2}{3} N^{3/2}$, yields
  \be
 I_{NUT}^{S^7} = \frac{ \sqrt2 \pi N^{3/2} }3   \,,
 \ee
which coincides with the ABJM free energy on the $S^3$ at the conformal point, as computed in \cite{Jafferis:2011zi}.

Our remaining task is now matching the gravitational on shell action of the set of Bolt solutions via a field theory computation, keeping in mind that these solutions are subject to the uplift constraint \eqref{quant_uplift} as well. We refer to \cite{Toldo:2017qsh} for the details of the large $N$ computation and limit ourselves here to stating the final result. For the ABJM theory, there is a saddle point solution for the large $N$ partition function which is:
\be \label{abjmms1} \log Z_{\cM_{g,p}}^{ABJM} = \frac{2\pi N^{3/2}}{3} \sqrt{2 k [m_1][m_2][m_3][m_4]} \bigg( -2 p + \sum_{i=1}^4 \frac{- p n_i +s_i + (g-1) r_i}{[m_i]} \bigg) \,, \ee
in function of the masses $m_i= [m_i] +n_i$, the flavor fluxes $s_i$, and finally the R-charges of the chiral multiplets $r_i$. The expression \eqref{abjmms1} is subject to the constraints $\sum_i [m_i]=1$,  $\sum_i m_i= \sum_i s_i =0$, $\sum_i r_i=2$.

There is moreover a shift symmetry which allows us to shift the $m_i$ by integers, therefore only the quantities $[m_i]$ and $-p m_i + s_i + (g-1) r_i$ are gauge invariant. To make contact with the dual setup, which is minimal gauged supergravity, it is natural to impose that
\be \label{mincond} [m_i] \;\;\; \text{and} \;\;\; {\mathfrak n}_i \equiv-p n_i +s_i +(g-1) r_i  \;\;\; \text{ are independent of}\; i \,.\ee
Following \cite{Benini:2015eyy}, we may interpret the fractional part of the masses, $[m_i]$, as fixing the asymptotic behavior of the scalar fields in the bulk\footnote{Notice that minimal gauged supergravity amounts to setting all scalars constant, in particular equal to their value in the vacuum (i.e. no radial flow). The value attained at the vacuum is independent of $i$.}.  Then ${\mathfrak n}_i$ can be attributed to the net magnetic charge felt by the $i$th chiral field. The condition \eqref{mincond} applied to the large $N$ expression of the partition function sets   $[m_i]=\frac{1}{4}$ and
$ \sum_i {\frak n}_i = p+2(g-1) $. If we set all the ${\mathfrak n}_i$ equal, then this can only be satisfied for ${\frak n}_i \in \mathbb{Z}$ if we impose:
\be \label{constraint1}  p+2 (g-1) = 0 \quad \text{mod} \quad 4 \ee
which is exactly the quantization condition required from the 11d uplift on $S^7$ to be well defined. Inserting these values into \eqref{abjmms1}, and taking into account the other companion solution of the Bethe Ansatz equation \cite{Toldo:2017qsh}, which accounts for the other branch of Bolt solutions, we find:

\be \label{fe1} \log Z_{\cM_{g,p}}^{ABJM} = \frac{\pi N^{3/2} \sqrt{2k} }{12} \big(p \pm 4 (g-1 ) \big) =I_{Bolt\pm}^{S^7} \,. \ee

\vspace{3mm}
\noindent For ABJM therefore we have matched separately the NUT and Bolts free energy with a field theory computation. Notice that for this model, the uplift condition \eqref{quant_uplift} does not allow for Bolt solutions with $p=1$ and $g=0$, namely with $S^3$ boundary. However, different Sasaki-Einstein truncations, arising as duals to more general 3d Chern-Simons models, can allow for multiple configurations with different topology to fill the same $S^3$ boundary. 
\hfill
\vspace{3mm}
\hfill
\begin{minipage}{0.6\textwidth} 
For instance, the truncation on the 7d manifold $M^{3,2}$ ($I=1$) with uplift condition \eqref{quant_uplift} allows for two Bolts and one NUT filling the squashed sphere boundary, with 
$$I_{Bolt_{-}} = \frac{5}{4} I_{NUT}\,, \qquad I_{Bolt_{+}} = \frac{3}{4} I_{NUT}\,,$$ The dual field theory computation however involves a chiral-like quiver, and its large $N$ limit is not under control.
\end{minipage}
\hfill
\begin{minipage}{0.37\textwidth}
\includegraphics[width=7.4cm]{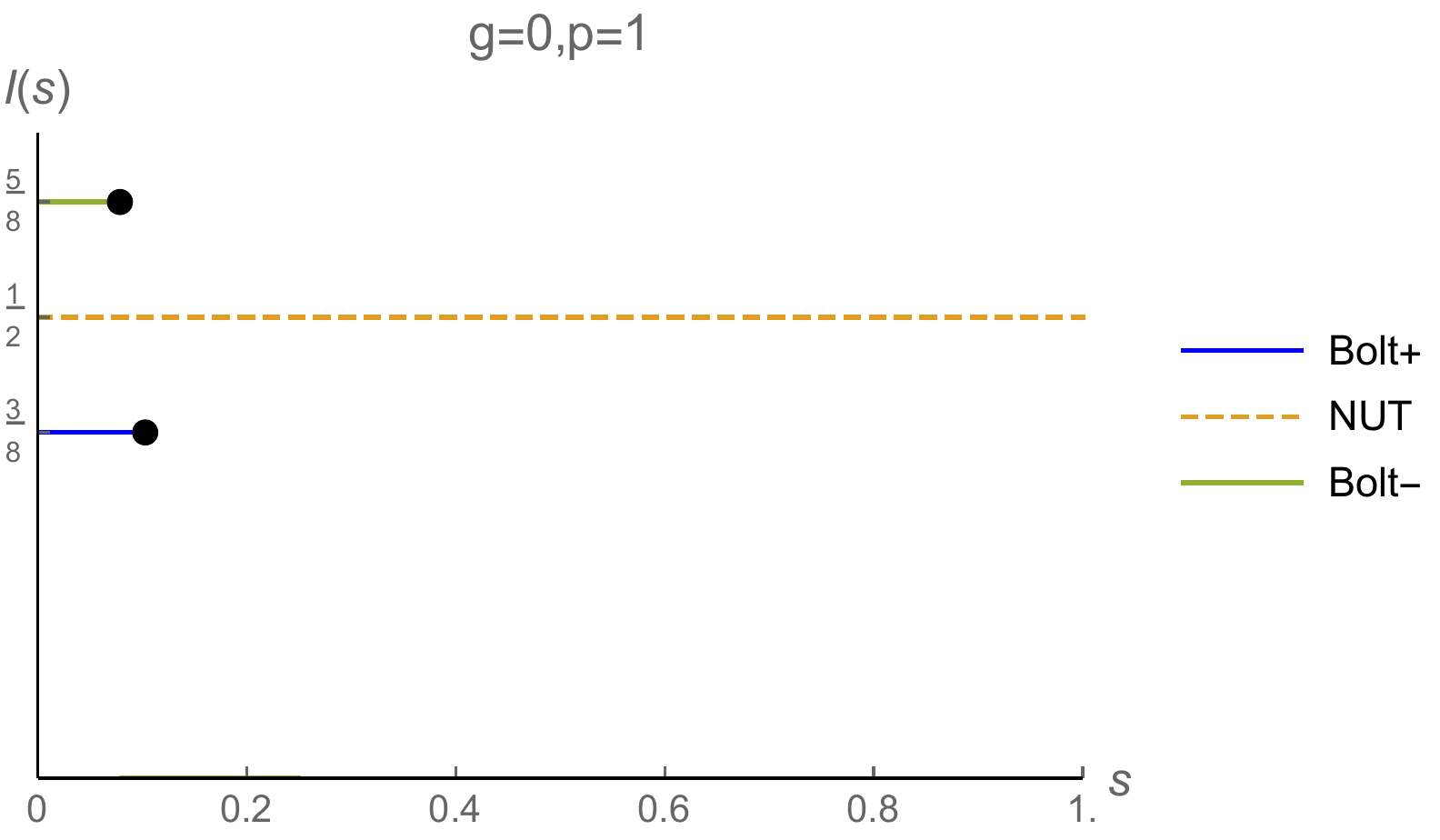}
\end{minipage}

\vspace{4mm}

 \noindent The figure shows the value of the on-shell action $I(s)$ for the three different solutions, which exist for certain intervals of squashing parameter $s$.  The uplift on the manifold $V^{5,2}=SO(5)/SO(3)$, instead, allows the Bolt$_-$ (denoted with the green color in the picture) as a regular solution: the large $N$ $\mathcal{M}_{g,p}$ partition function can be computed \cite{Toldo:2017qsh} and it indeed matches $I_{Bolt_{-}}$. This solution is a subleading saddle point of the path integral: the NUT solution always uplift and has lower free energy. The NUT on-shell action $I_{NUT} = \frac45 I_{Bolt_{-}}$ is matched by the field theory computation of \cite{Martelli:2011qj}, which makes use of a different eigenvalue distribution in the computation of $\log Z$. We refer the reader to the discussion on \cite{Toldo:2017qsh} for further details on this curious case. 

\section{Rotating black holes \label{rotating}}

In this section we describe the other class of gravity backgrounds mentioned in the introduction, related to the refinement by angular momentum: rotating BPS black holes.
For the solutions we consider, the value of the gauged R-symmetry magnetic flux (given by $\Iprod{G}{\Gamma}$ in the symplectically covariant notation we follow) distinguishes between two types of supersymmetry preserving asymptotics \cite{Hristov:2011ye}. When the flux vanishes we have the asymptotically AdS$_4$ solution, while the case when the flux is fixed to $-1$ is an example of a particular asymptotically locally AdS space that was dubbed "magnetic AdS" in \cite{Hristov:2011ye}. This dichotomy is well-understood on the AdS boundary for 3d supersymmetric theories \cite{Hristov:2013spa}, where either full superconformal symmetry is preserved ($\Iprod{G}{\Gamma}=0$) or there is only a partial supersymmetry via the topological twist ($\Iprod{G}{\Gamma}=-1$). 

Depending on the specific supergravity model, black hole solutions with a finite nonzero area of the event horizon might exist. The simplest solution is the supersymmetric electric Kerr-Newman-AdS$_4$ black hole in minimal gauged ${\cal N}=2$ supergravity, first analyzed in \cite{Kostelecky:1995ei,Caldarelli:1998hg}. More involved matter-coupled solution exist as well, and are summarized in Table 1. 

\begin{table}[h]
	\begin{center}
		\setlength{\tabcolsep}{3pt}
		\begin{tabular}{ c || c | c || c | c ||} \cline{2-5} & \multicolumn{2}{|c||}{AdS$_4\ (\Iprod{G}{\Gamma}=0)$} & \multicolumn{2}{|c||}{mAdS$_4\ (\Iprod{G}{\Gamma}=-1) $} \\ \hline
	\multicolumn{1}{|c||}{type}	 & gravity & $+$ matter & gravity & $+$ matter \\ \hline
		\multicolumn{1}{|c||}{$J=0$} & \cite{Romans:1991nq}\textasteriskcentered & \cite{Duff:1999gh}\textasteriskcentered & \cite{Romans:1991nq}\textasteriskcentered & \cite{Cacciatori:2009iz,Katmadas:2014faa,Halmagyi:2014qza} \\ \hline
		\multicolumn{1}{|c||}{$J\neq 0$} & \cite{Kostelecky:1995ei} & \cite{Cvetic:2005zi,Hristov:2019mqp} & \cite{Caldarelli:1998hg}\textasteriskcentered & \cite{Hristov:2018spe} \\ \hline
		\end{tabular}
	\end{center}
	\caption{Summary of known supersymmetric AdS$_4$ black holes with spherical topology. An asterisk denotes the absence of a regular horizon, i.e. a naked singularity.}
	\label{tab:sol}
\end{table}
The aim of \cite{Hristov:2018spe,Hristov:2019mqp} is to provide a systematic procedure to construct two classes of solution coupled to vector multiplets, with generic dyonic charges. Assuming that the associated scalar manifold is symmetric, we solve the BPS equations for any such model defined by an arbitrary gauging vector. From the point of view of holography, the main object of interest is the entropy function, which upon extremization with respect to a set of chemical potentials conjugate to the conserved charges gives the entropy of the newly discovered solutions. The Legendre transform of the entropy function presented above is expected to match the saddle point evaluation of the partition function of the holographically dual theory on (Euclidean) S$^1 \times$S$^2$. In the case of \cite{Hristov:2018spe}, this is the topologically twisted Witten index with angular momentum refinement, introduced in \cite{Benini:2015noa}. In the second case \cite{Cvetic:2005zi,Hristov:2019mqp}, the object is the generalized superconformal index \cite{Kim:2009wb,Imamura:2011su,Kapustin:2011jm}, whose large $N$ limit was recently found to capture the entropy of Kerr-Newman AdS$_4$ black holes \cite{Bobev:2019zmz,Benini:2019dyp,Nian:2019pxj} and matter coupled ones \cite{Choi:2019zpz}, the latter in  the Cardy limit.

\subsection{The real formulation of supergravity}\label{sec:formalism}
Our starting point is the action for abelian gauged ${\cal N} = 2$ supergravity with $n_V$ vector multiplets, with the same conventions as in \cite[Sec.~2]{Hristov:2018spe}. The bosonic fields are the metric $g_{\mu\nu}$, $(n_V+1)$ abelian gauge fields $A^I_\mu (I = 0, .., n_V)$  and $n_V$ complex scalars $z^i (i = 1, ..., n_V)$. The Lagrangian and supersymmetry transformation rules are uniquely specified by a choice of the so-called prepotential $F (X^I)$ and the symplectic vector of Fayet-Iliopoulos (FI) parameters $G = \{g^I, g_I \}$ defining the gauging. The BPS equations for solutions with a timelike Killing vector were conveniently written in \cite{Cacciatori:2008ek, Meessen:2012sr, Chimento:2015rra}, with a metric 
\begin{equation}\label{eq:metr-bps}
{\rm d} s^2_4 = -\e^{2 U} ({\rm d} t+\omega \? {\rm d}\phi)^2 + \e^{-2 U} {\rm d} s^2_3 \,,
\end{equation}
where ${\rm d} s^2_3$ is the metric of a three-dimensional base space, on which all quantities are defined. The gauge field strengths and their duals are packaged into a symplectic vector F$= (F_{\mu \nu}^{\Lambda}, G_{\mu \nu, \Lambda})$, which can be decomposed as
\be
\text{F} = d (\zeta (d t + \omega)) + \mathcal{F} = d (\zeta (d t + \omega)) + d \mathcal{A}
\ee
where $\zeta, \mathcal{F}, \mathcal{A}$ are symplectic vectors of timelike components and spatial field strengths and potentials, for both electric and magnetic gauge fields. We express the original complex scalars $z^i = X^i/X^0$ and scale factor $\e^{U}$ in terms of the symplectic section 
\begin{equation}
\e^{-2 U} \qv + \im \sv  = \{X^I, F_I \} \,,
\end{equation}
where $F_I \equiv \partial F/ \partial X^I$. Using the timelike isometry one may reduce the original four-dimensional action down to three dimensions, which leads to the so-called real formulation of special geometry \cite{Mohaupt:2011aa,Klemm:2012yg}. We choose to express everything in terms of ${\cal I}$ in our explicit ansatz, noting that one can further use 
\begin{equation}\label{eq:r-i}
 \qv = - \frac1{2 I_4(\sv)} I^\prime_4(\sv) = - \frac12 \e^{4 U} I^\prime_4(\sv)\,,
\end{equation}
In writing \eqref{eq:r-i} we already assumed that the special K\"{a}hler manifold parametrized by the scalar fields is a symmetric space, such that we can use the quartic invariant formalism reviewed in \cite[Sec.~2.2]{Hristov:2018spe}. The quartic form $I_4$ is invariant under symplectic transformations, while its derivative $I_4'$ is a symplectic vector and is therefore covariant. Moreover, the quartic invariant provides an explicit solution for the Hesse potential, thing that translates more explicitly in the following useful relation between the warp factor $U$ and the sections:
\be
e^{2U} = \sqrt{I_4(\mathcal{R})} \,.
\ee
We will be especially interested in the so-called $STU$ model, $
F= 2 i \sqrt{X^0 X^1 X^2 X^3}$, and purely electric gauging $G = \{0, g_I\}$, because the resulting Lagrangian can be embedded in 11d supergravity compactified on S$^7$ \cite{Duff:1999gh,Cvetic:1999xp}. 

\subsection{BPS equations and two different 3D base space ans\"{a}tze \label{2ansatz}}

The BPS equations for solutions of abelian gauged $\mathcal{N}=2$ supergravity with a timelike Killing vector were given in \cite{Cacciatori:2008ek, Meessen:2012sr, Chimento:2015rra}. We start from the BPS equations as summarized conveniently in the latter paper for the timelike class, for which the metric takes the form \eqref{eq:metr-bps}, as appropriate for black hole solutions. 

The BPS equations fix the gauge field strengths $\cF$ in terms of the scalars, so that the Maxwell equations and Bianchi identities take the form of a Poisson equation on the base metric, as 
\begin{equation}\label{BPS1}
d \cF = 
-d\left[ \star d \sv - 2\, e^{-4\?U}\,\Iprod{\star \hat{G}}{\qv}\,\qv + \? e^{-2\?U}\mathrm{J}\star\hat{G} \right] +d \omega\wedge \hat{G} 
= 0\,,
\end{equation}
where $\mathrm{J}$ is the scalar dependent complex structure, $\hat{G}$ for the model taken into consideration is the direct product of the vector of gaugings with a one-form. Introducing the vielbein $e^x$, with $x$, $y, ... =1,2,3$, for the three-dimensional base metric $ds^2_3$, $\hat{G}$  must satisfy the equation
\begin{equation}\label{BPS2}
 d e^x - \Iprod{\hat{G}}{\sv} \wedge e^x + \varepsilon^{xyz}\Iprod{\cal A}{\hat{G^y}}\wedge e^z =0 \,,
\end{equation}
where ${\cal A}$ denotes the spatial gauge potentials. The final BPS equation imposes that the rotation one-form $\omega$ must satisfy
\begin{equation}   \label{BPS3}
\star d\omega = \, \Iprod{d \sv}{\sv} - 2\? e^{-4\?U}\,\Iprod{\hat{G}}{\qv} = \, e^{-4\?U}\,\Iprod{\qv}{d \qv + 2 \?\hat{G} }\,,
\end{equation}
where in the second line we re-expressed the first term through the variable $\qv$. The conditions \eqref{BPS1}, \eqref{BPS2} and \eqref{BPS3} are sufficient to preserve supersymmetry.

The Poisson equation \eqref{BPS1} guarantees the local existence of the spatial gauge field strengths, $\cF$, which can be obtained by writing this equation as a total derivative. However, this is subtle in general due to the last term on the RHS, as it can be written as a total derivative in more than one way. In particular, invariance of $\cF$ following from \eqref{BPS1} under time reparametrizations, $\omega \rightarrow \omega + d \sigma$, for any function $\sigma$ on the base, requires that $\hat G$ be exact on a simply connected manifold. We write $  G = G\? d\rho $
for some function $\rho$, to be determined by \eqref{BPS2} once a particular base is chosen. The flow equations can then be simplified using the quartic invariant identities displayed for instance in Appendix A of \cite{Hristov:2018spe}, to find eventually algebraic BPS equations in terms of the combination $\sv$, allowing for simpler manipulation using techniques similar to \cite{Katmadas:2014faa, Halmagyi:2014qza}.

We use the following metric on the $3d$ base 
\begin{equation}
	{\rm d} s^2_3 = {\rm d} \rho^2 + \e^{2 \varphi} ({\rm d} x^2 + {\rm d} y^2)\ ,
\end{equation}
for a general function $\varphi(\rho, x, y)$. For stationary black hole solutions, one further assumes that $\partial/\partial y$ is also an isometry, leaving us with $\varphi(\rho, x)$. We find two classes of black hole solutions, depending on the separability of $\varphi$. The first one leads to the Cacciatori-Klemm-type solutions \cite{Cacciatori:2009iz,Katmadas:2014faa,Halmagyi:2014qza} \be
\e^{2 \varphi}_{CK} = \Phi(x)\ \e^{2 \psi(\rho)} \label{CKlike}
\ee
and their rotating generalizations \cite{Hristov:2018spe} which are described in Sec. \ref{twisted}. The second one leads to to Kerr-Newman type solutions \cite{Kostelecky:1995ei,Cvetic:2005zi,Hristov:2019mqp}, described in Sec. \ref{KNADS}, and in this case $\e^{2 \varphi}$ is separable in terms of new coordinates $q$ and $p$, such that
\begin{equation}\label{KNlike}
	\e^{2 \varphi} = Q(q) P(p)\ , \quad \rho = q\ p\ , \quad x = \alpha(q) + \beta(p)\ ,
\end{equation} 
with arbitrary functions $Q(q), P(p)$, while the functions $\alpha(q)$ and $\beta(p)$ are conventionally chosen as
$ \alpha^\prime(q) = - \frac{q}{Q(q)}\,,$ $ \beta^\prime(p) = \frac{p}{P(p)} $ in order to bring the base metric in the diagonal form
\begin{equation}\label{eq:3d-base}
 {\rm d} s^2_3 = \e^{2 \sigma} \left( \frac{{\rm d} p^2}{P(p)}  + \frac{{\rm d} q^2}{Q(q)} \right) + Q(q) P(p) {\rm d} y^2 \,,
\qquad 
 \e^{2 \sigma} \equiv q^2 P(p) + p^2 Q(q) \,.
\end{equation}
The standard form of the base metric for supersymmetric Kerr--Newman is reached upon setting $\{ q, p, y \} \sim \{r, \cos\theta, \phi \}$
where $r$ is a radial coordinate and $\theta$, $\phi$ are coordinates on a sphere. 

Before concluding the section, let us mention that while our  "ad hoc" choice for the two different three-dimensional bases successfully lead to new solutions, we cannot rule out that other BPS branches (for instance falling into the classes of \cite{Chow:2013gba,Gnecchi:2013mja}) might be found by using different ones. We limit however here to the two ans\"atze \eqref{CKlike},\eqref{KNlike} and in the next section we turn to the first class of solutions, which provide a rotating generalization of the topologically twisted magnetic black hole solutions of \cite{Cacciatori:2009iz}.

\subsection{Rotating topologically twisted BPS black holes: near horizon solution \label{twisted}}
Restricting to an attractor solution, which we expect to be topologically AdS$_2\times \Sigma$, we have to impose an appropriate scaling with respect to the radial coordinate for all the relevant fields. In particular, we take the function $\psi(r)$ in \eqref{CKlike} as
\begin{equation}\label{eq:psi-atrr}
 e^\psi = \vvv \, r \,,
\end{equation}
 with $\vvv$ a constant that physically gives the ratio between the scales of $\Sigma$ and AdS$_2$ on the horizon, so that the three-dimensional base metric becomes
 \begin{equation}\label{eq:base-bps-sph}
 ds^2_3 = dr^2 + \vvv^2\, r^2  \left(  \frac{d\theta^2}{\Delta(\theta)} + \?\Delta(\theta)\? f_\kappa^2(\theta)\, d\phi^2 \right) \, ,\quad 
f_\kappa(\theta) = \left\{ 
\begin{array}{ll} \sin\theta & \kappa = 1 \\ 1 & \kappa=0 \\ \sinh\theta & \kappa=-1 \end{array} \right.
\end{equation}
where $\kappa = 1$ for the spherical case, $\kappa = 0$ for the cylindrical, and $\kappa = -1$ for the hyperbolic case. We  parametrized the metric on $\Sigma$ in terms of a single function $\Delta(\theta)$, where $\theta$, $\phi$ are coordinates along the surface $\Sigma$. The direction $\phi$ corresponds to a compact isometry in all three cases. 

The conical structure of this ansatz implies the scaling behaviour
$ e^{-2U} = \frac{1}{r^2} e^{-2 u}$, $ \omega = \frac{1}{r}\omega_0$ for the remaining objects in the 4d metric, where $u$ and $\omega_0$ are a function and a one-form that depend only on the coordinates on $\Sigma$. The choice \eqref{eq:base-bps-sph} implies that the variables $\zeta$, $\qv$ and $\sv$ behave as
\begin{equation}\label{eq:scal-fields}
 \zeta = r\? \zeta_0 \,, \qquad  \qv = r\? \qv_0 \,, \qquad  \sv = \frac1r\? \sv_0 \,, 
\end{equation}
where $\zeta_0$, $\qv_0$ and $\sv_0$ are symplectic vectors that depend only on the coordinates on $\Sigma$ and are such that
$ e^{4 u} = I_4(\qv_0) = I_4(\sv_0)^{-1} $. Finally, one can show that with the choice \eqref{eq:psi-atrr}, the condition $\Iprod{G}{\sv_0} = 1 $ follows.

By the assumption that the scalar fields and $e^u$ do not depend on the radial coordinate near the horizon, this flow equation breaks up in components which $i)$ determine the dependence of the scalars along $\Sigma$, $ii)$ fix their constant parts in terms of the charges, directly generalizing the corresponding static attractor equation. This results in the following form for the warp factors appearing in the metric and the rescaled sections $ \sv_0$:

\begin{equation} \nonumber
\omega_0 =  - \frac{j}{\vvv}\? \left( 1  - \Iprod{{\cal H}_0}{I^\prime_4(G)}  \? j \? \cos \theta\ + I_4(G)\? j^2 \sin^2 \theta\ \right)\? \sin^2 \theta\ \?d\phi \,,
\end{equation}
\begin{equation}\nonumber
\Delta(\theta) =  1  - \Iprod{\cH_0}{I^\prime_4(G)}  \? j \? \cos \theta\ + I_4(G)\? j^2 \sin^2 \theta\ \,,
\end{equation}
\begin{equation}
 \vvv \sv_0  = {\cal H}_0 + j\? F_\kappa\? G\,, \qquad \vvv = \Iprod{G}{{\cal H}_0}\,, \qquad \Iprod{G}{\Gamma} = -1\,,
\end{equation}
for a constant symplectic vector, ${\cal H}_0$. We have defined $\partial_\theta F_\kappa \equiv - f_\kappa$. The final attractor equation, which relates the constant part of the scalars ${\cal H}_0$ to the electromagnetic charges $\Gamma$, becomes in the spherical case
\begin{equation}\label{eq:attr-fin}
  \Gamma = \frac{1}{4}\? I^\prime_4\left({\cal H}_0, {\cal H}_0, G \right) + \frac{1}{2}\? j^2\? I^\prime_4\left( G \right)\,.
\end{equation}
It generalizes the static attractor equation, which is obtained by setting $j=0$, and can be explicitly solved in a given model defined by a prepotential and a gauging vector $G$. 

\subsubsection{Full-flow solution}\label{sec:nearhor2}

Here we consider the full BPS flow for rotating black holes, interpolating between the attractor solution in the previous section and asymptotically locally AdS$_4$. Inspired by the known solutions in the static case \cite{Cacciatori:2009iz}, we take a metric of the form 
\begin{equation}\label{eq:metr-bps-full}
ds^2_4 = -e^{2U} \?(dt + \omega )^2 + e^{-2U}\?\left(  dr^2 + \frac{ e^{2\?\psi} }{\Delta(\theta)} d\theta^2 +  e^{2\?\psi}\?\Delta(\theta)\?\sin^2\!\theta\? d\phi^2 \right)  \,.
\end{equation}
where $U$ and $\psi$ depend on both radial and angular coordinate.  We introduce the combination 
\begin{equation}\label{eq:ItoH}
 e^\psi \? \sv = \cH (r) +  j \cos \theta\ G\ ,
\end{equation}

Here, we restrict for brevity on generalizing the simpler class of \cite{Katmadas:2014faa} to the rotating case, expecting that the most general solution of the static equations in \cite{Halmagyi:2014qza} can be also treated along the same lines. We therefore adopt the simple ansatz
\begin{equation}\label{eq:flow-ansatz}
\cH (r) = \cH_0 + \cH_\infty\ r  \,,   \qquad   \omega =  \left( \ominf(\theta)  - j \? \frac{\Delta(\theta)}{e^{\psi}} \right) \?\sin^2\!\theta\?d\phi \,,
\end{equation}
where the two constant symplectic vectors $ \cH_0, \cH_\infty$ and the function $\ominf(\theta)$ are to be determined. This ansatz reduces to the attractor solution of the previous section for $r\rightarrow 0$, while $\cH_\infty$ and $\ominf$ parametrize the asymptotic region. Inserting \eqref{eq:flow-ansatz} in the BPS equations leads to 
\begin{align}\label{eq:epsi-full}
 e^\psi = &\, \frac12\? \Iprod{G}{\cH_\infty}\?r^2 + \Iprod{G}{\cH_0}\?r  =  I_4(G)^{1/4}\?r^2 + \Iprod{G}{\cH_0}\?r  \,,
\end{align}
where we disregarded an additive integration constant and in the second equality we imposed that the $\cO(r^2)$ term is such that \eqref{eq:metr-bps-full} allows for AdS$_4$ asymptotics. 
We can solve this flow equation order by order in $r$ for the $d\theta \wedge d \phi$ terms.
Starting with the ${\cal O}(r^2)$ terms, we find that they are solved by fixing the vector $
\cH_\infty = \frac{1}{2}\? I_4(G)^{-3/4}\?I_4(G)^{\prime}\,,
$ in exactly the same way as in the static case \cite{Katmadas:2014faa}, and so on with the subleading terms recursively \cite{Hristov:2018spe} in order to find the full solution. The configuration has boundary metric
\begin{equation} \label{as_bound}
ds^2= r^2 \Delta(\theta) \left[- \frac{dt^2}{l^2} +\frac{d \theta^2}{\Delta(\theta)^2}+ \frac{\sin^2 \theta}{\Delta(\theta)} \left( d \phi +  \frac{j}{l^3} dt\right)^2  \right]\,,
\end{equation}
where $l = (I_4 (G))^{-1/4}$ sets the AdS$_4$ radius. The line element \eqref{as_bound} can be analytically continued to an Euclidean one by Wick-rotating time $t \rightarrow \im \tau $ and taking $j$ purely imaginary. This gives a conformal boundary of the form \eqref{refj_intro} as anticipated\footnote{The metric in square bracket is that of the Einstein space $\mathbb{R} \times S^2$ seen by a rotating frame of reference: a simple coordinate transformation  \cite{Hristov:2018spe} can bring the metric \eqref{as_bound} into the standard $\mathbb{R} \times S^2$ up to a conformal factor. In this form, locally the conformal boundary does not contain information about the chemical potential for angular momentum. However, as discussed at length in \cite{Cabo-Bizet:2018ehj}, the angular momentum chemical potential dictates the periodicity of the angular coordinates after one imposes regularity of the Euclideanized bulk solution. In other words, it is the global analysis that makes the role of the chemical potential manifest.}.
The Bekenstein-Hawking entropy $S_{BH}$ of the solution is
\begin{equation}\label{eq:BHentropy-horizon}
 S_{BH} = \frac{A}{4} = \pi\?\sqrt{I_4(\cH_0) - \left( 1 + I_4(G)\? j^2 \right)\? j^2} \,.
\end{equation}
The angular momentum $J$ can be computed via the Komar integral from the asymptotic fall offs of the full flow solution. This allows to solve \eqref{eq:attr-fin} for $j$ and $\cH_0$ in terms of $\cJ$ and the electromagnetic charges $\Gamma$, in order to obtain the entropy formula in terms of conserved charges. This is the expression that should be reproduced by the large $ N$ limit of the topologically twisted Witten index with angular momentum refinement \cite{Benini:2015noa,Closset:2018ghr}. The evaluation of the large $N$ limit  of such quantity is to date still an open issue. 

Finally, the expression for the entropy function as was first found \cite{Hristov:2018spe} is quite cumbersome and not particularly illuminating -- it simplifies drastically only for a special class of models that does not exhibit AdS asymptotics. Let us mention that \cite{Hosseini:2019iad} provided a convenient expression for the entropy functional, inspired by the factorized structure of the known supersymmetric partition functions, which makes the attractor mechanism manifest: the values of the scalar  fields  at  the  horizon are identified with  the  critical  points  of  the  entropy  functional. It would be interesting to retrieve such expression from a field theory computation.

\subsection{Rotating electric Kerr-Newman-AdS$_4$ black holes: near-horizon solution \label{KNADS}}

This section deals with the second class of solutions, those dubbed as "Kerr-Newman- type" in Sec. \ref{2ansatz}. We start the resolution of the BPS equation by making the choice $\hat{G} = G\?{\rm d}(q\?p)$, and using the redefinition of the symplectic section $\cH = \e^{2 \sigma} \?\sv\,$ and $I_4(\cH) = \e^{8 \sigma}\e^{- 4U}$. This change of variable brings the BPS equations to a form that can be solved in terms of polynomial ans\"{a}tze for the variables $\e^{2 \sigma}$ and $\cH$. We first want to solve the BPS equations \eqref{BPS1}-\eqref{BPS3} near the horizon, therefore we impose an ansatz compatible with an AdS$_2$ factor in the geometry. We choose here $q \equiv r$ and the function $Q(r)$ as $ Q(r) = R_0^2 r^2$, where $r$ is a radial coordinate and $R_0$ is a constant, so that $\e^{2 \sigma}$ is also separable: 
$ \e^{2 \sigma} = r^2 \e^{2 \sigma_0} \,,$ and  $  \e^{2 \sigma_0} =  P(p) + R_0^2 p^2 $.
The conical structure of the base space implies the scaling behaviour
\begin{equation}\label{eq:eU-attr}
 \e^{-2  U} = \frac{1}{r^2} \e^{-2 U_0}\,, \quad \omega = \frac{1}{r} \omega_0 \,, \quad  \cH = r\, \cH_0\,,
\end{equation}
for the components of the metric and the scalars, where the functions $U_0, \omega_0$ and $\cH_0$ depend only on $p$. With this ansatz, we are left with solving the BPS equations in order to determine the dependence on the coordinate $p$, i.e.\ along the sphere. From \eqref{BPS1} and \eqref{BPS3} we  obtain 
\begin{equation}\label{eq:hor-omega}
 \omega_0 = j\? P(p)\? \e^{-2 \sigma_0} = j \frac{P(p)}{ P(p) + R_0^2 p^2 }\,,
\end{equation}
where $j$ is a constant to be fixed in due course. 
The remaining equations can be written in terms of $\e^{2\?\sigma_0}$, the symplectic vector $\cH_0$ and its contractions with the vector $G$. A polynomial ansatz for $\cH_0$
\begin{equation}\label{eq:H0-ansatz-hor}
 \cH_0 = \cC_3\? p^3 + \cC_2 \? p^2 + \cC_1 \? p + \cC_0 \,,
\end{equation}
allows to integrate the BPS equation for $\e^{\sigma_0}$ as
\begin{eqnarray}\label{eq:solhorsigma}
 \e^{2\?\sigma_0} = &\, \frac{1}{2}\?\Iprod{G}{\cC_3}\? p^4 + \frac{2}{3}\?\Iprod{G}{\cC_2}\? p^3 +\Iprod{G}{\cC_1}\? p^2 + 2\?\Iprod{G}{\cC_0}\? p + \Xi^{-1}\,,
\end{eqnarray}
with $\Xi$ an integration constant. The remaining components of the BPS equations are then solved order by order in $p$, expressing the parameters $\mathcal{C}_i$ in function of a single constant symplectic vector ${\cal C}$ \cite{Hristov:2019mqp}. The rotation parameter $j$ is fixed in terms of ${\cal C}$ from
$
 j = -\frac{1}{2\,\Xi}\? \Iprod{G}{I_4^\prime(\cC)} \,.
$
Finally, the gauge field strengths are given by
\begin{equation}\label{eq:fieldstr}
	{\cal F} = R_0^2\? {\rm d} \left( \e^{-2\?\sigma_0} p\? \left( \cH_0 - j\?  p^{2}\? G \right) \? {\rm d} y  \right) \,.
\end{equation}
Restricting to spherical horizon topology and absence of NUT charge, we further get
\be \label{eq:R0value}
 \Iprod{G}{\cC} = 0\ , \qquad \Iprod{I_4^\prime(G)}{I_4^\prime(\cC)} = 0\,,   \qquad 	\Xi\? R_0^2 = 1+I_4(G)\ I_4(\cC) +\frac{1}{4} I_4 (\cC,\cC,G,G)\ .
\ee
Upon the redefinition $p = \cos \theta\ ,$ $ y = \phi\ $ we arrive at the familiar expression  $P(\theta) = \frac{\sin^2 \theta}{\Xi}\ \left(1 - \frac{a^2}{l^2} \cos^2 \theta \right)\,,$ with $ \,\Xi \equiv (1 - \frac{a^2}{l^2}) $, $ a \equiv  \frac{\sqrt{I_4(\cC)}}{l} $. The resulting charge vector is computed through \eqref{eq:fieldstr}, as
\begin{equation}\label{eq:attractoreq}
 \Gamma  \equiv \frac{1}{4\?\pi}\?\int \cF  = \frac{1}{\Xi} \left(\cC + \frac{1}{8}\?I_4^\prime(I_4^\prime(\cC), G, G) \right) \,.
\end{equation}
This constitutes the attractor equation, through which the scalars and metric functions encoded in the vector $\cC$ can be solved for in terms of the charges $\Gamma = \{P^I, Q_I\}$. This leads to the constraint on the magnetic flux of the R-symmetry $
	\Iprod{G}{\Gamma} = 0 $, while the Bekenstein-Hawking entropy yields
\begin{eqnarray}\label{eq:Bek-Haw-entropy}
	S_{BH} =& \frac{\pi}{\Xi}\left(  \Xi\? R_0^2 \? I_4(\cC) - \frac{1}{4} \Iprod{G}{I_4'(\cC)}^2 \right)^{1/2} \ ,
\end{eqnarray}
where $R_0$ is given by \eqref{eq:R0value}. Notice that solving the constraint \eqref{eq:R0value} for instance for the $STU$ model leaves us with up to six free parameters: the black hole is characterized by four independent electric and two independent magnetic charges.

\subsubsection{Full flow black hole solution}\label{sec:fullflow}
In constructing full black hole geometries we use $q=r$ as a radial variable that running from the horizon to the asymptotic AdS$_4$ spacetime. It is natural to extend the near-horizon behaviour of $\cH$ in \eqref{eq:eU-attr} to the more general polynomial ansatz 
\begin{equation}\label{eq:H-flow-ans}
 \cH = r\? \left( \cH_0(p) + \left( \cH_1^{(0)} + \cH_1^{(1)} p \right)\?r + \cH_2^{(1)} p \? r^2 \right)\,,
\end{equation}
where all vectors $\cH_{1}^{(0,1)}$ and $\cH_{2}^{(1)}$ are constant, while $\cH_0$ is the horizon value. To maintain AdS$_4$ asymptotics, the function $\e^{2\?\sigma}$ in \eqref{eq:3d-base} needs to be quartic in $r$. Given this, we have
\begin{equation}
 \e^{2\?\sigma} = r^2\?\left( \e^{2\?\sigma_0}  + \frac{2}{3}\?\Iprod{G}{\cH_1^{(1)}\?r + \cH_2^{(1)}\?r^2 }\?p^2  \right)\ ,
\end{equation}
where we disregarded integration constants and imposed $\Iprod{G}{\cH_1^{(0)}}  = 0\,,$ in order to keep the structure assumed in the near-horizon geometry. We also make the following ansatz for the rotation form $\omega$ 
\begin{equation}\label{eq:omega-full}
 \omega = \e^{-2\?\sigma}\left( \mu\?Q(r) + j\?r\?P(p)  \right) - \mu\,,
\end{equation}
where $j$ was fixed already on the horizon, while $\mu$ is a second integration constant. With a bit of work,  from the BPS equations \eqref{BPS1}-\eqref{BPS3} one sees that the solution is fully fixed in terms of the vectors $\cC$ and $G$, whose values of $\cH_i$ are reported in \cite{Hristov:2019mqp}. The metric function $Q$ becomes
\begin{equation}\label{eq:Qofr}
	Q(r) = \frac{r^2}{\Xi} \left( \Xi\?R_0^2  + l\? \Iprod{\cC}{I_4'(G)}\? r + \frac{r^2}{l^2} \right)\ ,
\end{equation}
where the first term in the bracket also depends explicitly on the vectors $\cC$ and $G$ via \eqref{eq:R0value}. The solutions found here asymptote to AdS$_4$, with conformal boundary of the form \eqref{refj_intro}. The mass and angular momentum satisfy the familiar BPS bound $M =  \frac{|{\cal J}|}{l} + \frac{1}{\sqrt{2}} \,  \big| \sum_{I=1}^4 Q_I  \big| $.

Finally, the Bekenstein-Hawking entropy can be obtained upon extremization of the so called "entropy function" $S$, with respect to a set of chemical potentials conjugate to the conserved charges. We report here the expression of the entropy function:
\begin{align} \label{eq:entropyfn}
	S = -2\? \frac{F (X)}{\omega} - F_I (X) P^I + X^I Q_I    + \frac{\omega}{2}\?\left( {\cal J} + P^I Q_I \right) + \lambda\? (2\? g_I X^I - \omega - 2 \pi i)\ ,
\end{align}
where the $X^I$ are conjugate to the electric charges $Q_I$, $\omega$ is conjugate to ${\cal J}$, $g_I$ are the FI gauging parameters, and the prepotential $F(X)$ and its derivatives $F_I(X)$. The Lagrange multiplier $\lambda$ imposes a constraint among the chemical potentials such that upon extremization with respect to the independent set of $\omega, X^I$ one recovers the entropy $S_{BH}$. In the absence of magnetic charges $P^I$, the above entropy function was  introduced in \cite{Choi:2018fdc} and further elaborated in \cite{Cassani:2019mms}, based on the previously known black hole solutions in \cite{Cvetic:2005zi}. We confirm the conjecture of \cite{Choi:2018fdc} for the full $STU$ model with electric charges and its extension for a new BPS solution to the $X^0 X^1$ model including a magnetic charge.  The Legendre transform of the entropy function should match the large $N$ limit of the partition function of the dual field theory on S$^1 \times$S$^2$ with angular momentum refinement \cite{Kim:2009wb,Imamura:2011su,Kapustin:2011jm}. Evidence for this, in the Cardy limit, was recently provided in \cite{Choi:2019zpz}, where among others it was shown that the superconformal index exhibits the correct $N^{3/2}$ scaling, necessary to reproduce the correct multiplicity of black hole states, upon taking complex fugacities.


\begin{thebibliography}{99}

  %\cite{Toldo:2017qsh}
\bibitem{Toldo:2017qsh} 
  C.~Toldo and B.~Willett,
  %``Partition functions on 3d circle bundles and their gravity duals,''
  JHEP {\bf 1805}, 116 (2018)
    [arXiv:1712.08861 [hep-th]].
  %%CITATION = doi:10.1007/JHEP05(2018)116;%%


\bibitem{Hristov:2018spe} 
  K.~Hristov, S.~Katmadas and C.~Toldo,
  %``Rotating attractors and BPS black holes in $AdS_4$,''
  JHEP {\bf 1901}, 199 (2019)
  [arXiv:1811.00292 [hep-th]].
 

\bibitem{Hristov:2019mqp} 
  K.~Hristov, S.~Katmadas and C.~Toldo,
  %``Matter-coupled supersymmetric Kerr-Newman-AdS$_4$ black holes,''
  Phys.\ Rev.\ D {\bf 100}, no. 6, 066016 (2019)
  [arXiv:1907.05192 [hep-th]].
  %%CITATION = doi:10.1103/PhysRevD.100.066016;%%
  
  %\cite{Strominger:1996sh}
\bibitem{Strominger:1996sh} 
  A.~Strominger and C.~Vafa,
  %``Microscopic origin of the Bekenstein-Hawking entropy,''
  Phys.\ Lett.\ B {\bf 379}, 99 (1996)
  [hep-th/9601029].
  %%CITATION = doi:10.1016/0370-2693(96)00345-0;%%
    
    \bibitem{Maldacena:1997de} 
  J.~M.~Maldacena, A.~Strominger and E.~Witten,
  %``Black hole entropy in M theory,''
  JHEP {\bf 9712}, 002 (1997)
  [hep-th/9711053].
    
    
 %\cite{Benini:2015eyy}
\bibitem{Benini:2015eyy} 
  F.~Benini, K.~Hristov and A.~Zaffaroni,
  %``Black hole microstates in AdS$_{4}$ from supersymmetric localization,''
  JHEP {\bf 1605}, 054 (2016)
  [arXiv:1511.04085 [hep-th]].
  %%CITATION = doi:10.1007/JHEP05(2016)054;%%
  
  %\cite{Aharony:2008ug}
\bibitem{Aharony:2008ug}
  O.~Aharony, O.~Bergman, D.~L.~Jafferis and J.~Maldacena,
  %``N=6 superconformal Chern-Simons-matter theories, M2-branes and their gravity duals,''
  JHEP {\bf 0810} (2008) 091
  [arXiv:0806.1218 [hep-th]].

 %\cite{Zaffaroni:2019dhb}
\bibitem{Zaffaroni:2019dhb} 
  A.~Zaffaroni,
  %``Lectures on AdS Black Holes, Holography and Localization,''
  arXiv:1902.07176 [hep-th].
  %%CITATION = ARXIV:1902.07176;%%
 
 %\cite{Closset:2017zgf}
\bibitem{Closset:2017zgf} 
  C.~Closset, H.~Kim and B.~Willett,
  %``Supersymmetric partition functions and the three-dimensional A-twist,''
  JHEP {\bf 1703}, 074 (2017)
  [arXiv:1701.03171 [hep-th]].
  %%CITATION = doi:10.1007/JHEP03(2017)074;%%
         
   %\cite{Jafferis:2011zi}
\bibitem{Jafferis:2011zi} 
  D.~L.~Jafferis, I.~R.~Klebanov, S.~S.~Pufu and B.~R.~Safdi,
  %``Towards the F-Theorem: N=2 Field Theories on the Three-Sphere,''
  JHEP {\bf 1106}, 102 (2011)
  [arXiv:1103.1181 [hep-th]].
  
  \bibitem{Casini:2011kv} 
  H.~Casini, M.~Huerta and R.~C.~Myers,
  %``Towards a derivation of holographic entanglement entropy,''
  JHEP {\bf 1105}, 036 (2011)
  [arXiv:1102.0440 [hep-th]].
  %%
 
  %\cite{Alday:2012au}
\bibitem{Alday:2012au} 
  L.~F.~Alday, M.~Fluder and J.~Sparks,
  %``The Large N limit of M2-branes on Lens spaces,''
  JHEP {\bf 1210}, 057 (2012)
  [arXiv:1204.1280 [hep-th]].
  

 \bibitem{Benini:2015noa} 
  F.~Benini and A.~Zaffaroni,
  %``A topologically twisted index for three-dimensional supersymmetric theories,''
  JHEP {\bf 1507}, 127 (2015)
  [arXiv:1504.03698 [hep-th]].
  %%CITATION = doi:10.1007/JHEP07(2015)127;%%
  
    %\cite{Martelli:2012sz}
\bibitem{Martelli:2012sz} 
  D.~Martelli, A.~Passias and J.~Sparks,
  %``The supersymmetric NUTs and bolts of holography,''
  Nucl.\ Phys.\ B {\bf 876}, 810 (2013)
  [arXiv:1212.4618 [hep-th]].
 

   %\cite{Closset:2018ghr}
\bibitem{Closset:2018ghr} 
  C.~Closset, H.~Kim and B.~Willett,
  %``Seifert fibering operators in 3d $\mathcal{N}=2$ theories,''
  JHEP {\bf 1811}, 004 (2018)
  [arXiv:1807.02328 [hep-th]].
  %%CITATION = doi:10.1007/JHEP11(2018)004;%%
   
   
%\cite{Klemm:2011xw}
\bibitem{Klemm:2011xw} 
  D.~Klemm,
  %``Rotating BPS black holes in matter-coupled $AdS_4$ supergravity,''
  JHEP {\bf 1107}, 019 (2011)
  [arXiv:1103.4699 [hep-th]].
  %%CITATION = doi:10.1007/JHEP07(2011)019;%%
   
   \bibitem{Kim:2009wb} 
  S.~Kim,
  %``The Complete superconformal index for N=6 Chern-Simons theory,''
  Nucl.\ Phys.\ B {\bf 821}, 241 (2009)
  Erratum: [Nucl.\ Phys.\ B {\bf 864}, 884 (2012)]
  [arXiv:0903.4172 [hep-th]].
  %%CITATION = doi:10.1016/j.nuclphysb.2012.07.015, 10.1016/j.nuclphysb.2009.06.025;%%
  %2
  
  %\cite{Imamura:2011su}
\bibitem{Imamura:2011su} 
  Y.~Imamura and S.~Yokoyama,
  %``Index for three dimensional superconformal field theories with general R-charge assignments,''
  JHEP {\bf 1104}, 007 (2011)
  [arXiv:1101.0557 [hep-th]].
  %%CITATION = doi:10.1007/JHEP04(2011)007;%%
  
  %\cite{Kapustin:2011jm}
\bibitem{Kapustin:2011jm} 
  A.~Kapustin and B.~Willett,
  %``Generalized Superconformal Index for Three Dimensional Field Theories,''
  arXiv:1106.2484 [hep-th].
  %%CITATION = ARXIV:1106.2484;%%
  %114 c

   %\cite{Kostelecky:1995ei}
\bibitem{Kostelecky:1995ei} 
  V.~A.~Kostelecky and M.~J.~Perry,
  %``Solitonic black holes in gauged N=2 supergravity,''
  Phys.\ Lett.\ B {\bf 371}, 191 (1996)
  [hep-th/9512222].
  %%CITATION = doi:10.1016/0370-2693(95)01607-4;%%
 
 %\cite{Cvetic:2005zi}
\bibitem{Cvetic:2005zi} 
  M.~Cvetic, G.~W.~Gibbons, H.~Lu and C.~N.~Pope,
  %``Rotating black holes in gauged supergravities: Thermodynamics, supersymmetric limits, topological solitons and time machines,''
  hep-th/0504080.
  %%CITATION = HEP-TH/0504080;%%
  

 %\cite{Chamblin:1998pz}
\bibitem{Chamblin:1998pz} 
  A.~Chamblin, R.~Emparan, C.~V.~Johnson and R.~C.~Myers,
  %``Large N phases, gravitational instantons and the nuts and bolts of AdS holography,''
  Phys.\ Rev.\ D {\bf 59}, 064010 (1999)
  [hep-th/9808177].
  %%CITATION = doi:10.1103/PhysRevD.59.064010;%%
  

 %\cite{Plebanski:1976gy}
\bibitem{Plebanski:1976gy} 
  J.~F.~Plebanski and M.~Demianski,
  %``Rotating, charged, and uniformly accelerating mass in general relativity,''
  Annals Phys.\  {\bf 98}, 98 (1976).
  
    %\cite{Martelli:2013aqa}
\bibitem{Martelli:2013aqa} 
  D.~Martelli and A.~Passias,
  %``The gravity dual of supersymmetric gauge theories on a two-parameter deformed three-sphere,''
  Nucl.\ Phys.\ B {\bf 877}, 51 (2013)
  [arXiv:1306.3893 [hep-th]].
  
  
   %\cite{Benini:2016hjo}
\bibitem{Benini:2016hjo} 
  F.~Benini and A.~Zaffaroni,
  %``Supersymmetric partition functions on Riemann surfaces,''
  Proc.\ Symp.\ Pure Math.\  {\bf 96}, 13 (2017)
  [arXiv:1605.06120 [hep-th]].
 
  %\cite{AlonsoAlberca:2000cs}
\bibitem{AlonsoAlberca:2000cs} 
  N.~Alonso-Alberca, P.~Meessen and T.~Ortin,
  %``Supersymmetry of topological Kerr-Newman-Taub-NUT-AdS space-times,''
  Class.\ Quant.\ Grav.\  {\bf 17}, 2783 (2000)
  [hep-th/0003071].
  
  \bibitem{Martelli:2011fw} 
  D.~Martelli and J.~Sparks,
  %``The gravity dual of supersymmetric gauge theories on a biaxially squashed three-sphere,''
  Nucl.\ Phys.\ B {\bf 866}, 72 (2013)
  [arXiv:1111.6930 [hep-th]].
  

 %\cite{Gauntlett:2007ma}
\bibitem{Gauntlett:2007ma} 
  J.~P.~Gauntlett and O.~Varela,
  %``Consistent Kaluza-Klein reductions for general supersymmetric AdS solutions,''
  Phys.\ Rev.\ D {\bf 76}, 126007 (2007)
  [arXiv:0707.2315 [hep-th]].
  
  %\cite{Skenderis:2002wp}
\bibitem{Skenderis:2002wp} 
  K.~Skenderis,
  %``Lecture notes on holographic renormalization,''
  Class.Quant.Grav.\  {\bf 19},5849(2002)
  [hep-th/0209067].
  %%CITATION = doi:10.1088/0264-9381/19/22/306;%%
  
   %\cite{Caldarelli:1998hg}
\bibitem{Caldarelli:1998hg} 
  M.~M.~Caldarelli and D.~Klemm,
  %``Supersymmetry of Anti-de Sitter black holes,''
  Nucl.\ Phys.\ B {\bf 545}, 434 (1999)
  [hep-th/9808097].
  
  %\cite{Azzurli:2017kxo}
\bibitem{Azzurli:2017kxo} 
  F.~Azzurli, N.~Bobev, P.~M.~Crichigno, V.~S.~Min and A.~Zaffaroni,
  %``A universal counting of black hole microstates in AdS$_{4}$,''
  JHEP {\bf 1802}, 054 (2018)
  [arXiv:1707.04257 [hep-th]].
  %%CITATION = doi:10.1007/JHEP02(2018)054;%%

\bibitem{BenettiGenolini:2019jdz} 
  P.~Benetti Genolini, J.~M.~Perez Ipina and J.~Sparks,
  %``Localization of the action in AdS/CFT,''
  JHEP {\bf 1910}, 252 (2019)
  [arXiv:1906.11249 [hep-th]].
    
  %\cite{Hama:2011ea}
\bibitem{Hama:2011ea} 
  N.~Hama, K.~Hosomichi and S.~Lee,
  %``SUSY Gauge Theories on Squashed Three-Spheres,''
  JHEP {\bf 1105}, 014 (2011)
  [arXiv:1102.4716 [hep-th]].
  %%CITATION = doi:10.1007/JHEP05(2011)014;%%
  %

\bibitem{Martelli:2011qj} 
  D.~Martelli and J.~Sparks,
  %``The large N limit of quiver matrix models and Sasaki-Einstein manifolds,''
  Phys.\ Rev.\ D {\bf 84}, 046008 (2011)
  [arXiv:1102.5289 [hep-th]].
  %%CITATION = doi:10.1103/PhysRevD.84.046008;%%
  
 %\cite{Hristov:2011ye}
\bibitem{Hristov:2011ye} 
  K.~Hristov, C.~Toldo and S.~Vandoren,
  %``On BPS bounds in D=4 N=2 gauged supergravity,''
  JHEP {\bf 1112}, 014 (2011)
  [arXiv:1110.2688 [hep-th]].
  %%CITATION = doi:10.1007/JHEP12(2011)014;%%
  %
  
  %\cite{Hristov:2013spa}
\bibitem{Hristov:2013spa} 
  K.~Hristov, A.~Tomasiello and A.~Zaffaroni,
  %``Supersymmetry on Three-dimensional Lorentzian Curved Spaces and Black Hole Holography,''
  JHEP {\bf 1305}, 057 (2013)
  [arXiv:1302.5228 [hep-th]].
  
  
  
 %\cite{Romans:1991nq}
\bibitem{Romans:1991nq} 
  L.~J.~Romans,
  %``Supersymmetric, cold and lukewarm black holes in cosmological Einstein-Maxwell theory,''
  Nucl.\ Phys.\ B {\bf 383}, 395 (1992)
  [hep-th/9203018].
  %%CITATION = doi:10.1016/0550-3213(92)90684-4;%%

 %\cite{Duff:1999gh}
\bibitem{Duff:1999gh} 
  M.~J.~Duff and J.~T.~Liu,
  %``Anti-de Sitter black holes in gauged N = 8 supergravity,''
  Nucl.\ Phys.\ B {\bf 554}, 237 (1999)
  [hep-th/9901149].
  %%CITATION = doi:10.1016/S0550-3213(99)00299-0;%%
 
    

 %\cite{Cacciatori:2009iz}
\bibitem{Cacciatori:2009iz} 
  S.~L.~Cacciatori and D.~Klemm,
  %``Supersymmetric AdS(4) black holes and attractors,''
  JHEP {\bf 1001}, 085 (2010)
  [arXiv:0911.4926 [hep-th]].
  %%CITATION = doi:10.1007/JHEP01(2010)085;%%
  %146 citations counted in INSPIRE as of 23 Dec 2019
 
     \bibitem{Katmadas:2014faa} 
  S.~Katmadas,
  %``Static BPS black holes in U(1) gauged supergravity,''
  JHEP {\bf 1409}, 027 (2014)
  [arXiv:1405.4901 [hep-th]].
  %%CITATION = doi:10.1007/JHEP09(2014)027;%%
 
 
 %\cite{Halmagyi:2014qza}
\bibitem{Halmagyi:2014qza} 
  N.~Halmagyi,
  %``Static BPS black holes in AdS$_{4}$ with general dyonic charges,''
  JHEP {\bf 1503}, 032 (2015)
  [arXiv:1408.2831 [hep-th]].
  %%CITATION = doi:10.1007/JHEP03(2015)032;%%

 %\cite{Bobev:2019zmz}
\bibitem{Bobev:2019zmz} 
  N.~Bobev and P.~M.~Crichigno,
  %``Universal spinning black holes and theories of class $ \mathcal{R} $,''
  JHEP {\bf 1912}, 054 (2019)
  [arXiv:1909.05873 [hep-th]].
  %%CITATION = doi:10.1007/JHEP12(2019)054;%%
  
  %\cite{Benini:2019dyp}
\bibitem{Benini:2019dyp} 
  F.~Benini, D.~Gang and L.~A.~Pando Zayas,
  %``Rotating Black Hole Entropy from M5 Branes,''
  arXiv:1909.11612 [hep-th].
  %%CITATION = ARXIV:1909.11612;%%
  
  %\cite{Nian:2019pxj}
\bibitem{Nian:2019pxj} 
  J.~Nian and L.~A.~Pando Zayas,
  %``Microscopic Entropy of Rotating Electrically Charged AdS$_4$ Black Holes from Field Theory Localization,''
  arXiv:1909.07943 [hep-th].
  %%CITATION = ARXIV:1909.07943;%%
  
  %\cite{Choi:2019zpz}
\bibitem{Choi:2019zpz} 
  S.~Choi, C.~Hwang and S.~Kim,
  %``Quantum vortices, M2-branes and black holes,''
  arXiv:1908.02470 [hep-th].
  %%CITATION = ARXIV:1908.02470;%%

   
 %\cite{Cacciatori:2008ek}
\bibitem{Cacciatori:2008ek} 
  S.~L.~Cacciatori, D.~Klemm, D.~S.~Mansi and E.~Zorzan,
  %``All timelike supersymmetric solutions of N=2, D=4 gauged supergravity coupled to abelian vector multiplets,''
  JHEP {\bf 0805}, 097 (2008)
  [arXiv:0804.0009 [hep-th]].
  %%CITATION = doi:10.1088/1126-6708/2008/05/097;%%
  %6
  
  %\cite{Meessen:2012sr}
\bibitem{Meessen:2012sr} 
  P.~Meessen and T.~Ortin,
  %``Supersymmetric solutions to gauged N=2 d=4 sugra: the full timelike shebang,''
  Nucl.\ Phys.\ B {\bf 863}, 65 (2012)
  [arXiv:1204.0493 [hep-th]].
  %%CITATION = doi:10.1016/j.nuclphysb.2012.05.023;%%
  %42 citations counted in INSPIRE as of 23 Dec 2019
  
  %\cite{Chimento:2015rra}
\bibitem{Chimento:2015rra} 
  S.~Chimento, D.~Klemm and N.~Petri,
  %``Supersymmetric black holes and attractors in gauged supergravity with hypermultiplets,''
  JHEP {\bf 1506}, 150 (2015)
  [arXiv:1503.09055 [hep-th]].
  %%CITATION = doi:10.1007/JHEP06(2015)150;%%
  %21 citations counted in INSPIRE as of 23 Dec 2019
  
  \bibitem{Mohaupt:2011aa} 
  T.~Mohaupt and O.~Vaughan,
  %``The Hesse potential, the c-map and black hole solutions,''
  JHEP {\bf 1207}, 163 (2012)
  [arXiv:1112.2876 [hep-th]].
  
  
  \bibitem{Klemm:2012yg} 
  D.~Klemm and O.~Vaughan,
  %``Nonextremal black holes in gauged supergravity and the real formulation of special geometry,''
  JHEP {\bf 1301}, 053 (2013)
  [arXiv:1207.2679 [hep-th]].
  
   %\cite{Cvetic:1999xp}
\bibitem{Cvetic:1999xp} 
  M.~Cvetic {\it et al.},
  %``Embedding AdS black holes in ten-dimensions and eleven-dimensions,''
  Nucl.\ Phys.\ B {\bf 558}, 96 (1999)
  [hep-th/9903214].
  %%CITATION = doi:10.1016/S0550-3213(99)00419-8;%%
  %4
  
   %\cite{Chow:2013gba}
\bibitem{Chow:2013gba} 
  D.~D.~K.~Chow and G.~Compere,
  %``Dyonic AdS black holes in maximal gauged supergravity,''
  Phys.\ Rev.\ D {\bf 89}, no. 6, 065003 (2014)
  [arXiv:1311.1204 [hep-th]].
  %%CITATION = doi:10.1103/PhysRevD.89.065003;%%

\bibitem{Gnecchi:2013mja} 
  A.~Gnecchi, K.~Hristov, D.~Klemm, C.~Toldo and O.~Vaughan,
  %``Rotating black holes in 4d gauged supergravity,''
  JHEP {\bf 1401}, 127 (2014)
  [arXiv:1311.1795 [hep-th]].
  %%CITATION = doi:10.1007/JHEP01(2014)127;%%

\bibitem{Cabo-Bizet:2018ehj} 
  A.~Cabo-Bizet, D.~Cassani, D.~Martelli and S.~Murthy,
  %``Microscopic origin of the Bekenstein-Hawking entropy of supersymmetric AdS$_{5}$ black holes,''
  JHEP {\bf 1910}, 062 (2019)
  [arXiv:1810.11442 [hep-th]].

% \cite{Hosseini:2019iad}
\bibitem{Hosseini:2019iad} 
  S.~M.~Hosseini, K.~Hristov and A.~Zaffaroni,
  %``Gluing gravitational blocks for AdS black holes,''
  arXiv:1909.10550 [hep-th].
  %%CITATION = ARXIV:1909.10550;%%
  
  %\cite{Choi:2018fdc}
\bibitem{Choi:2018fdc} 
  S.~Choi, C.~Hwang, S.~Kim and J.~Nahmgoong,
  %``Entropy functions of BPS black holes in AdS$_4$ and AdS$_6$,''
  arXiv:1811.02158 [hep-th].
  %%CITATION = ARXIV:1811.02158;%%
  %19 citations counted in INSPIRE as of 24 Dec 2019

%\cite{Cassani:2019mms}
\bibitem{Cassani:2019mms} 
  D.~Cassani and L.~Papini,
  %``The BPS limit of rotating AdS black hole thermodynamics,''
  JHEP {\bf 1909},079 (2019)
[arXiv:1906.10148].
  %%CITATION = doi:10.1007/JHEP09(2019)079;%%
 

 
\end{thebibliography}
\end{document}